\documentclass[a4paper]{PoS}

\usepackage{bbold}

\newcommand{\cyi}{Computation-based Science and Technology
  Research Center, The Cyprus Institute, 20 Konstantinou Kavafi Str.,
  2121 Nicosia, Cyprus}

\newcommand{\ucy}{Department of Physics, University of Cyprus,
  P.O. Box 20537, 1678 Nicosia, Cyprus}

\newcommand{\desy}{NIC, DESY, Platanenallee 6, D-15738 Zeuthen,
  Germany}

\newcommand{\temple}{Department of Physics, Temple University,
  Philadelphia, PA}

\newcommand{\anl}{Physics Division, Argonne National Laboratory,
  Argonne, IL}
  
\newcommand{\Dlr}{\buildrel \leftrightarrow \over D\raise-1pt\hbox{}}

\title{Investigating volume effects for N$_f$=2 twisted clover fermions at the physical point}

\ShortTitle{Volume effects for twisted clover fermions}

\author{Constantia Alexandrou \\\ucy\, and \\\cyi\\  E-mail:
  \email{alexand@ucy.ac.cy}}
  
\author{Simone Bacchio \\\ucy\\ E-mail: 
    \email{s.bacchio@hpc-leap.eu}}

\author{Martha Constantinou, Dean Howarth  \\\temple \\ E-mail: \email{marthac@temple.edu}, 
    \email{tug88731@temple.edu}}

\author{\speaker{Colin Lauer} \\\temple\, and \\\anl\\ E-mail: \email{colin.j.lauer@temple.edu}}

\author{Kyriakos Hadjiyiannakou,
  Giannis Koutsou \\\cyi\\  E-mail:
  \email{k.hadjiyiannakou@cyi.ac.cy},
  \email{g.koutsou@cyi.ac.cy}}

\author{Karl Jansen \\\desy\\  E-mail:
  \email{karl.jansen@desy.de}}

\abstract{In this work we present preliminary results on the nucleon axial and tensor charges, the quark momentum fraction and the first moment of the helicity distribution. The simulations have been performed using two $N_f{=}2$ and  $N_f{=}2{+}1{+}1$ ensembles. Both ensembles have a twisted mass action with a clover term with the quark masses fixed at their physical values (physical point) and volumes of $64^3{\times}128$. The extracted quantities are compared with results from an existing $N_f{=}2$ ($48^3{\times}96$) ensemble. This comparison allows one to address volume and quenching effects.}

\FullConference{The 36th Annual International Symposium on Lattice Field Theory - LATTICE2018\\
		22-28 July, 2018\\
		Michigan State University, East Lansing, Michigan, USA.}

\begin{document}

\section{Introduction}{
One of the most important efforts of particle and nuclear physics is to understand the structure of the nucleon from first principles. Despite being studied for many years, there are still several unknown aspects on the nucleon structure, such as the origin of its mass, the charged radii, and the distribution of its spin among its constituents. Lattice QCD (LQCD) offers an ideal method for \textit{ab initio} calculations which can be used to study the properties of fundamental particles numerically. The lattice formulation is a non-perturbative tool that allows the study of phenomena at the hadronic scale, where perturbation theory fails. 

Parton distribution functions (PDFs) are important quantities that give insights on the nucleon structure. For example, to leading twist, they represent how the probability density of finding a specific parton in a hadron depends on the hadron's momentum and spin. PDFs are in the light-cone frame so they cannot be calculated in Euclidean space. A way to access PDFs from lattice QCD is to calculate moments of PDFs which are related to the original PDF through the operator product expansion. These moments can also be calculated from deep inelastic scattering (DIS) experiments through phenomenological analysis which can be compared to the LQCD results \cite{Abdel-Rehim}. 

In the framework of this work we compute nucleon matrix elements of a variety of local bilinear operators (with up to one covariant derivative). This includes the scalar, vector, axial and tensor charges, as well as, the one-derivative vector, axial and tensor. In this proceedings we present results only for the nucleon axial and tensor charges, and the first moments of the unpolarized and helicity distribution functions. It is crucial to extract reliable estimates of the aforementioned quantities from lattice QCD as they can serve as benchmark of the lattice techniques (e.g., the axial charge), and even provide input in the analyses of experimental data (e.g., the tensor charge~\cite{Lin:2017stx}). In particular, the axial charge $g_A$ gives information about the chiral structure of a particle and its value for the nucleon at the chiral limit is used as an input for chiral effective theories. It has been thoroughly studied in chiral effective theories \cite{chi-eff} and is well measured experimentally in $\beta$-decay experiments. In addition, $g_A$ is an ideal candidate for a benchmark quantity because it can be calculated directly at zero momentum transfer, unlike the anomalous magnetic moment of the nucleon, for example, which must be extrapolated from finite momentum transfer calculations. 
The tensor charge $g_T$ is an interesting quantity that may have implications for physics beyond the Standard Model. It is necessary for setting bounds on novel tensor interactions in ultra-cold neutron decays which are beyond the standard model \cite{gT}. Unlike $g_A$, its value is not well known experimentally with only limits on its value coming from radiative pion decay  $\pi\rightarrow e\nu \gamma$. There are experiments at Jefferson Lab which use polarized $^3$He/Proton with the goal of increasing the experimental accuracy of $g_T$ by an order of magnitude \cite{jlab}, making a theoretical calculation especially timely.}

\section{Calculation and Simulation Details}{
\subsection{Matrix elements}

\begin{figure}
\vspace*{-0.5cm}
\hspace{.5cm}
\begin{minipage}{.45\linewidth}
    \vspace{1.25cm}
    \includegraphics[scale=0.35]{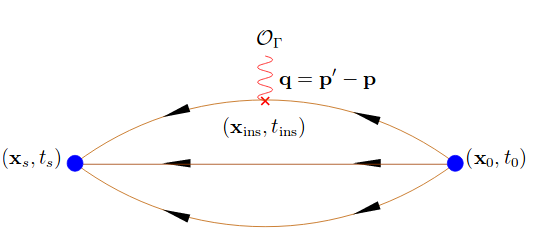}
\end{minipage}
\hspace{.5cm}
\begin{minipage}{.45\linewidth}
    \includegraphics[scale=0.35]{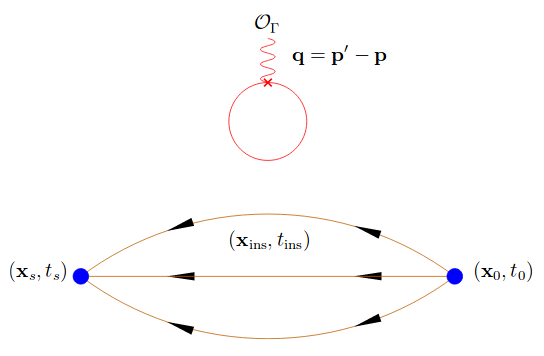}
\end{minipage}
\caption{Diagrams of connected (left) and disconnected (right) contributions to the three-point functions.}
\label{fig:3ptDiagrams}
\end{figure}
\vspace{1mm}

The calculation is based on nucleon matrix elements $\langle N(p)|{\cal O}_\Gamma|N(p)\rangle$, where the inserted operator, ${\cal O}_\Gamma$, is a local bilinear operator. To obtain these matrix elements, we calculate the three-point functions which are represented by the diagrams in Figure \ref{fig:3ptDiagrams}. In the connected diagram (left) the current couples with the quark fields of the nucleon interpolating field, while in the disconnected diagram (right) the quarks of the current for a closed loop and interact with the nucleon via gluon exchange. In this study, we focus on isovector quantities that receive contributions only from the connected diagram (the disconnected contributions cancel out up to cut-off effects). For $g_A$ and $g_T$, ${\cal O}_\Gamma$ is the ultra-local axial-vector and tensor operators, respectively: \\[0.5ex]
\begin{equation}
    \displaystyle{\cal O}^\mu_{A^a} = \bar{q}\gamma_5\gamma^\mu\frac{\tau^a}{2} q\>, \quad
    {\cal O}^{\mu\nu}_{T^a} =\bar{q}\sigma^{\mu\nu}\frac{\tau^a}{2}q\>.
\end{equation}
where $\bar{q}=(\bar{u},\bar{d})$. For $\langle x\rangle_q$ and $\langle x\rangle_{\Delta q}$, the three-point functions are calculated using the one-derivative vector and axial-vector operators that is: \\[0.5ex]
\begin{equation}
    {\cal O}_{V^a}^{\mu\nu} = \bar{q}\gamma^{\{\mu} \Dlr^{\nu\}} \frac{\tau^a}{2}q, \quad
    {\cal O}_{A^a}^{\mu\nu} = \bar{q}\gamma^{\{\mu} \Dlr^{\nu\}} \gamma_5 \frac{\tau^a}{2}q\,.
\end{equation}
The curly brackets represent a symmetrization over indice pairs and a subtraction of the trace. The presence of the derivative in the operators leads to reduction in the signal-to-noise ratio and larger statistics is required to extract $\langle x\rangle_q$ and $\langle x\rangle_{\Delta q}$ at the same statistical accuracy as the charges. The matrix elements of the ultra-local operators at zero momentum transfer give the nucleon charges ($g_A{\equiv}G_A(0)$ and $g_T{\equiv}A_{T10}(0)$) via the continuum decomposition:
\begin{equation}
    \langle N(p,s')|\mathcal{O}^\mu_{A}|N(p,s)\rangle = i \bar{u}_N(p,s') \Bigl[\frac{1}{2}G_A(0)\gamma^\mu\gamma_5\Bigr] u_N(p,s)\,,
    \label{eq:OpA}
\end{equation}
\begin{equation}
    \langle N(p,s')|\mathcal{O}^{\mu\nu}_{T}|N(p,s)\rangle = \bar{u}_N(p,s') \Bigl[\frac{1}{2} A_{T10}(0)\sigma^{\mu\nu} \Bigr] u_N(p,s)\,.\label{eq:OpT}
\end{equation}
Equivalently, the matrix elements for the one-derivative operators  at zero momentum transfer lead to:
\begin{equation}
    \langle N(p,s^\prime)| {\cal O}_{V}^{\mu\nu}|N(p,s)\rangle =
  \bar{u}_N(p,s^\prime)\Bigl[
    \frac{1}{2}\langle x\rangle_q \gamma^{\{\mu}p^{\nu\}}
    \Bigr] u_N(p,s)\,,
    \label{eq:oneD_V}
\end{equation}
\begin{equation}
    \langle N(p,s^\prime)| {\cal O}_{A}^{\mu\nu}|N(p,s)\rangle = i
  \bar u_N(p,s^\prime)\Bigl[
    \frac{1}{2} \langle x\rangle_{\Delta q}\gamma^{\{\mu}p^{\nu\}}\gamma^5
    \Bigr] u_N(p,s).
    \label{eq:oneD_A}
\end{equation}
For simplicity, we use the generic symbol $q$ to represent the quark combination in \ref{eq:oneD_V} and \ref{eq:oneD_A} but, in this proceedings, we will only consider the isovector case $q=u-d$.

\subsection{Lattice Setup}

The results presented in these proceedings are obtained from three different ensembles of the twisted-clover action
$$ S_F\left[\chi,\overline{\chi},U \right]= a^4\sum_x  \overline{\chi}(x)\left(D_W[U] + m_{\rm cr} + i \mu_l \gamma_5\tau^3 - \frac{1}{4}c_{\rm SW}\sigma^{\mu\nu}\mathcal{F}^{\mu\nu}[U] \right) \chi(x), $$
with quarks tuned to maximal twist. This action gives an automatic ${\cal O}(a)$-improvement and does not require any operator improvement, simplifying the renormalization procedure \cite{tm}. Table \ref{tab:simParams} shows the parameters used for each ensemble, listed from the smallest to the largest volume. The pion mass of each ensemble is near its physical value, thus avoiding a chiral extrapolation that may lead to uncontrolled uncertainties. The statistics are up to 24,000 for cB211.72.64 and 12,656 for cA2.90.64 (for their highest separation).
{\small
\renewcommand{\arraystretch}{1.2}
\renewcommand{\tabcolsep}{18pt}
\begin{table}[ht]
\centering
\begin{tabular}{c|c|c|c|c|c}
    \hline\hline
    \multicolumn{6}{c}{Parameters}\\
    \hline
    & $N_f$ & \hspace*{-0.2cm}Volume\hspace*{-0.2cm} & a (fm)  & \hspace*{-0.4cm}$m_\pi$ (MeV)\hspace*{-0.4cm} &  \hspace*{-0.4cm} $t_s/a$ \hspace*{-0.4cm} \\
        \hline
    \hspace*{-0.4cm}    cA2.90.48\hspace*{-0.4cm}    & \hspace*{-0.4cm}$2$\hspace*{-0.4cm} & \hspace*{-0.2cm}$48^3{\times}96$\hspace*{-0.2cm} & $0.094$  & $130$ & 10, 12, 14\\
    \hspace*{-0.4cm}    
    cB211.72.64 \cite{Nf211}\hspace*{-0.4cm} & \hspace*{-0.4cm}$2{+}1{+}1$\hspace*{-0.4cm} & \hspace*{-0.2cm}$64^3{\times}128$\hspace*{-0.2cm} & $0.082$  & $137$ & 12, 14, 16, 18, 20  \\
    \hspace*{-0.4cm}    
    cA2.90.64\hspace*{-0.4cm}    & \hspace*{-0.4cm}$2$\hspace*{-0.4cm} & \hspace*{-0.2cm}$64^3{\times}128$\hspace*{-0.2cm} & $0.094$ & $130$ & 12, 14, 16 \\\hline\hline
\end{tabular}
\caption{Parameters for the ensembles used in this work, and source-sink separation $t_s/a$.}
\label{tab:simParams}
\end{table}

\noindent
In order to extract the matrix elements we calculate the two- and three-point functions given by
\begin{equation}
    G_{\rm 2pt}({\bf 0}, t_s) = \sum_{{\bf x}_s}\Gamma^4_{\beta\alpha}\langle J_\alpha({\bf x}_s,t_s)\bar{J}_\beta({\bf x}_0,t_0)\rangle
\end{equation}
\begin{equation}
    G^{\mu_1,...,\mu_n}_{\rm 3pt}(\Gamma^\nu, {\bf p},  t_s, t_{\rm i}) =
  \sum_{{\bf x}_s,{\bf x}_{\rm i}}\> e^{-i({\bf x}_s-{\bf x}_0)\cdot{\bf p}}\>
  \Gamma^\nu_{\beta\alpha}\langle J_\alpha({\bf x}_s,t_s) {\cal O}_\Gamma^{\mu_1,...,\mu_n}
 ({\bf x}_{\rm i}, t_{\rm i})\bar{J}_\beta({\bf x}_0,t_0)\rangle,
\end{equation}
respectively, where $x_0$, $x_i$, and $x_s$ are the insertion, sink, and source coordinates. $\Gamma^\nu$ is the projection matrix 
$    \Gamma^4 = \frac{1}{4}(\mathbb{1}+\gamma_4), \  \Gamma^k = \Gamma^4i\gamma_5\gamma_k$, and the proton interpolation operators are
$    J_{\alpha}(x) = \epsilon^{abc}u_{\alpha}^a(x)[u^{\top b}(x) C\gamma_5 d^c(x)]$, where a, b, and c are color component indices. We then form the ratio:
\begin{equation}
    R(\Gamma^\nu,t_s,t_i)=
\frac{G_{\rm 3pt}(\Gamma^\nu,{\bf 0},t_s,t_i) }{G_{\rm 2pt}({\bf 0}, t_s)}
\label{eq:ratio}
\end{equation}
which is simplified due to the zero momentum transfer. The information on the nucleon charges can be extracted from the fits of the ratio in Eq.~(\ref{eq:ratio}) using different techniques. We focus here on a single- ({\it plateau}) and {\it two-state} fits in order to examine ground state dominance. Results on the summation method are not shown here as they carry large statistical uncertainties leading to inconclusive results.

\medskip
\noindent
{\it Plateau method:} In this method, the current insertion time ($t_i$) dependence of the ratio is fitted to a constant assuming a single-state dominance. The fit includes values of $t_i$ far from $t_o$ and $t_s$, so that contamination due to excited states is suppressed. The excited state effects are studied by varying $t_s$ and convergences is expected at large $t_s$. 

\medskip
\noindent
{\it Two-state method:} Instead of fitting the ratio to a constant value, alternatively one fits including the first excited state. In this case, the two- and three-point functions are written as
\begin{eqnarray}
    G_{2pt}(\vec{p},t_s) &=& c_0(\vec{p}) e^{-E_0(\vec{p}) t_s} + c_1(\vec{p}) e^{-E_1(\vec{p}) t_s}, \\[2ex]
    G_{3pt}(\vec{p}\,',\vec{p},t_s,t_i) &=& A_{00}(\vec{p}\,',\vec{p}) e^{-E_0(\vec{p}\,')(t_s-t_i)-E_0(\vec{p})t_i} + A_{01}(\vec{p}\,',\vec{p}) e^{-E_0(\vec{p}\,')(t_s-t_i)-E_1(\vec{p})t_i} \nonumber \\
&+& A_{10}(\vec{p}\,',\vec{p}) e^{-E_1(\vec{p}\,')(t_s-t_i)-E_0(\vec{p})t_i} + A_{11}(\vec{p}\,',\vec{p}) e^{-E_1(\vec{p}\,')(t_s-t_i\
)-E_1(\vec{p})t_i},
\label{Eq:Thrp_tsf}
\end{eqnarray}
where $E_0(\vec{p})$ and $E_1(\vec{p})$ are the ground and first excited state with total momentum $\vec{p}$. The two- and three-point functions are fitted simultaneously which involves twelve fitting parameters . The matrix element of interest $\mathcal{M}$ is then
${\cal M}=\frac{A_{00}(\vec{p}\,',\vec{p})}{\sqrt{c_0(\vec{p}\,') c_0(\vec{p})}}\,$,
and here we focus on zero momentum transfer for which $A_{10} {=} A_{01}$.

\section{Results}
\subsection{Nucleon Charges}

In Fig.~\ref{fig:gA} (left) we show the ratio of Eq.~(\ref{eq:ratio}) for $g_A$ as a function of $t_i$ using the ensemble cB211.72.64 (see Table~\ref{tab:simParams}). For clarity purposes we only show the two largest separations $t_s$ for the plateau (red circles and blue triangles), while the two-state fit is shown with a green band using all possible separations. We find perfect agreement between the shown data within statistical uncertainties. In the right plot, we collect the final values of $g_A$ for all ensembles (cA2.90.48 (red circles), cB211.72.64 (blue triangles), and cA2.90.64 (green square)) as a function of source-sink separation. With the blue horizontal band we show the two-state fit curve for cB211.72.64 and the curve's convergent values. The corresponding two-state fit for cA2.90.64 will be performed in the near future once statistics are increased together with higher separation $t_s$.

\begin{figure}
    \vspace*{-.3cm}
    \hspace*{-.65cm}
    \includegraphics[scale=0.29]{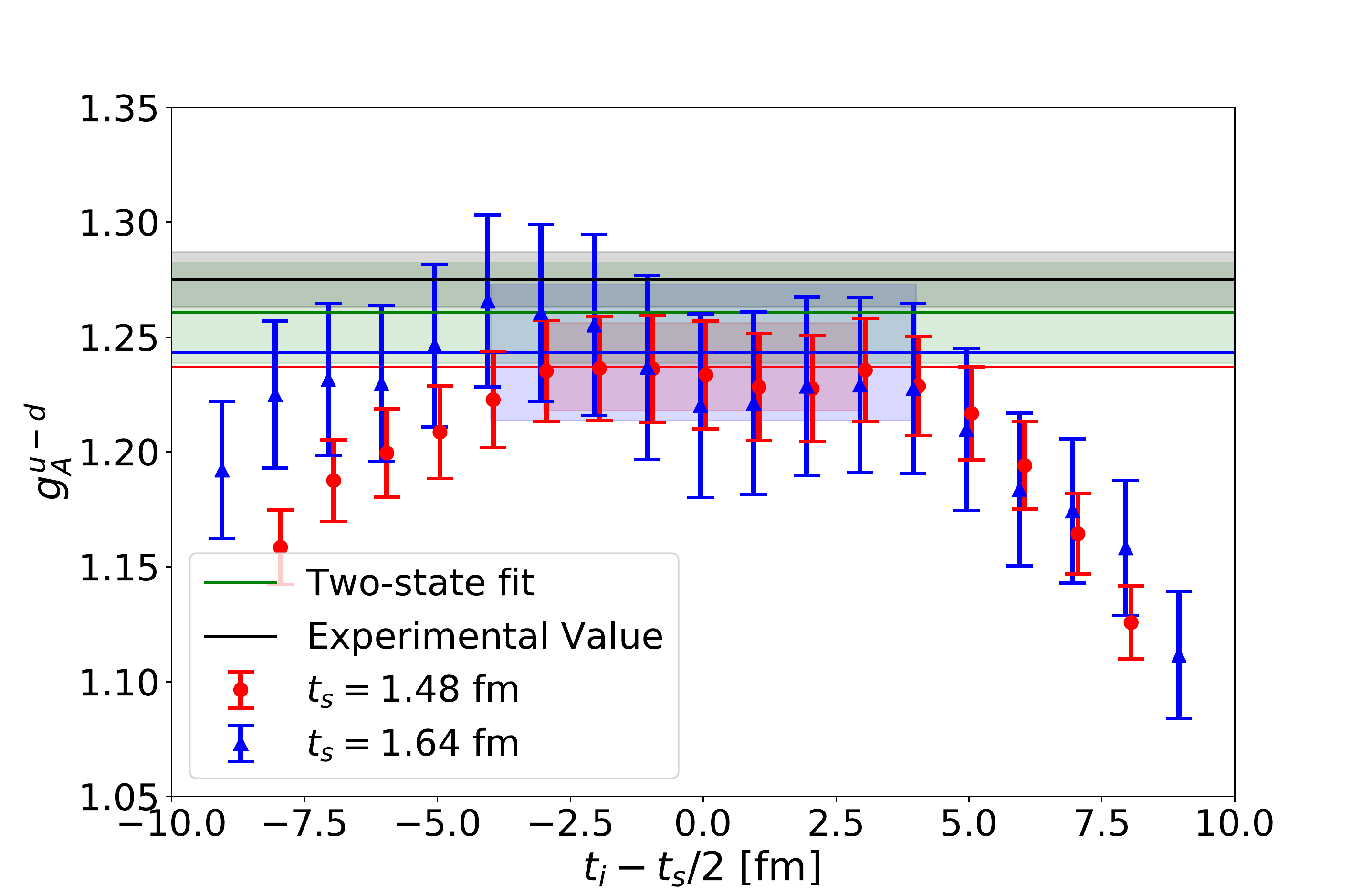}
    \hspace*{-.75cm}
    \includegraphics[scale=0.29]{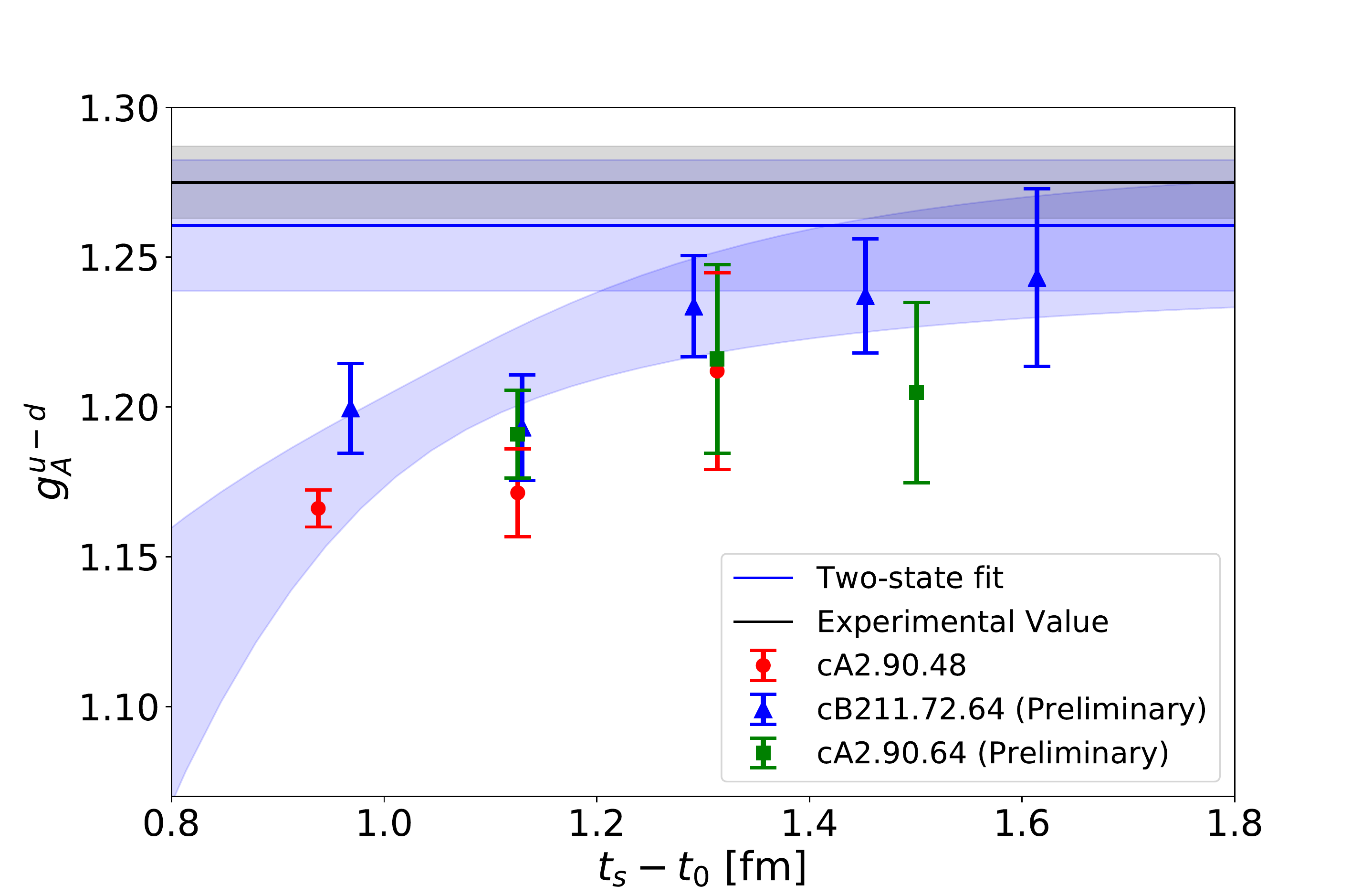}
    \caption{Left: the ratio of $g_A$ for cB211.72.64 and $t_s{=}18a$ (red circles) and $t_s{=}20$ (blue triangles). The green horizontal band is the two-state fit value. Right: plateau values for $g_A$ for cA2.90.48 (red circles), cB211.72.64 (blue triangles), and cA2.90.64 (green squares). The two-state fit curve and its convergent value for cB211.72.64 is also included. In both plots, the black horizontal band is the global fit of experimental data \cite{wp} and the errors show the statistical errors.
    }
    \label{fig:gA}

\end{figure}

\noindent
In the right panel of Fig.~\ref{fig:gA}, it can be seen that $g_A$ rises as $t_s$ increases for cA2.90.48 and cB211.72.64 which shows that excited state contamination is affecting the values at lower $t_s$. Although the values agree between ensembles, it is not apparent that the cA2.90.64 data are rising with increased $t_s$, and thus it is not yet conclusive if there are volume or quenching effects, as cA2.90.48 and cA2.90.64 only share two data points. New data are currently being produced at higher source-sink separations as well as higher statistics for existing points. This will reveal whether this behavior is observed because excited states are not being suppressed as quickly as for the other ensembles or simply because statistical noise is obscuring the increase. The data point at $t_s\sim 1.6$ fm is in agreement with the experimental value and the two-state fit value for cB211.72.64 is in a very good agreement with experiment. 

The results for $g_T$ are presented in Fig.~\ref{fig:gT}, in the same format as the right plot of of Fig.~\ref{fig:gA}. Unlike the case of $g_A$, $g_T$ decreases for larger $t_s$ values, showing effects due to excited state contamination. However, we do not find strong volume or quenching effects for $g_T$ at these lattice volumes, as all three ensembles agree within statistical uncertainty.
\vspace*{-0.375cm}
\begin{figure}
 \centering        
        \includegraphics[scale=0.265]{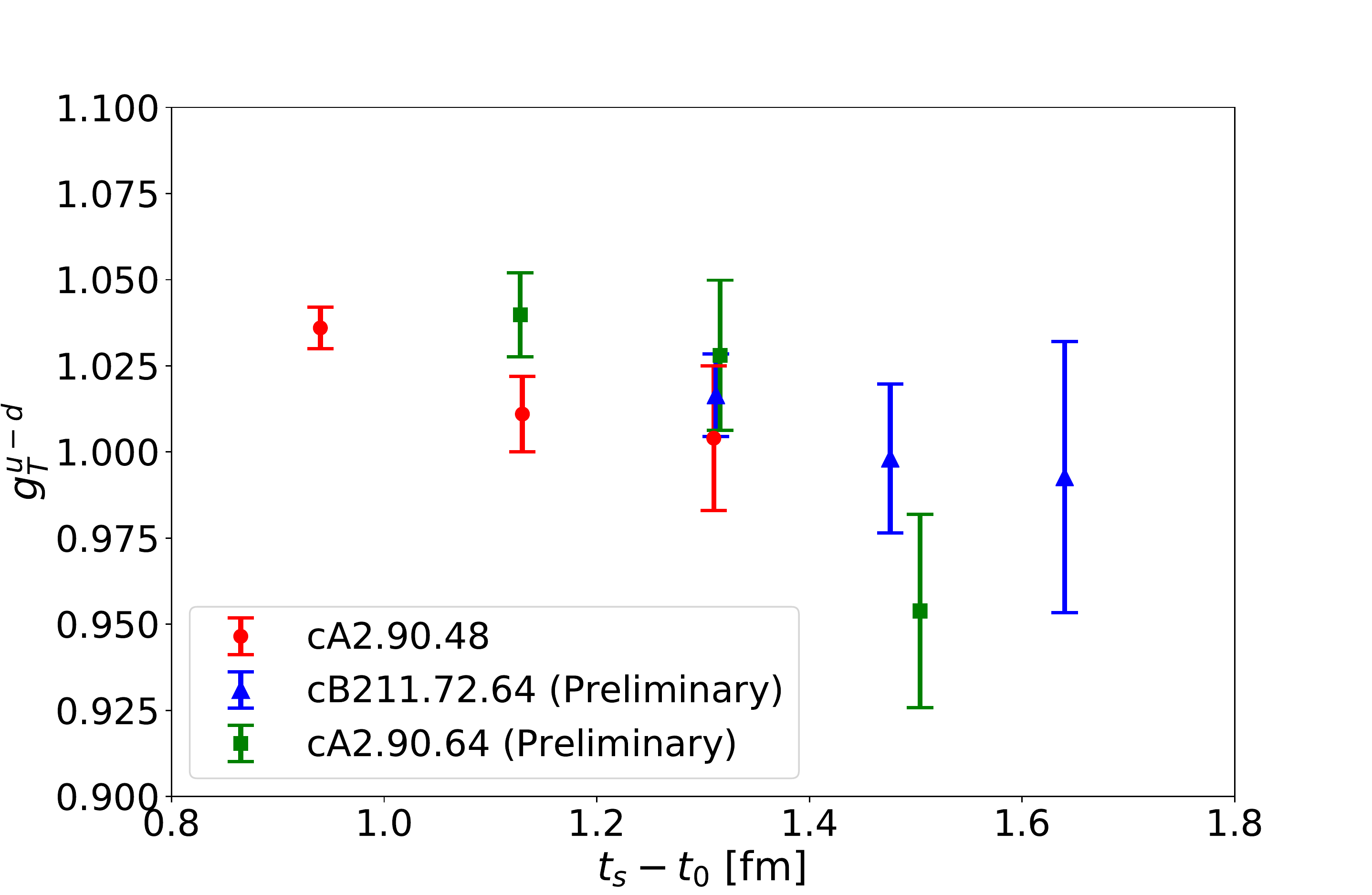}
        \vspace*{-0.2cm}
        \caption{Plateau values for $g_T$. Notation on the ensembles is the same as in the right plot of Fig.~\ref{fig:gA}.   }
        \label{fig:gT}
\end{figure}

\subsection{First moments of unpolarized and helicity PDFs}
Fig.~\ref{fig:PDFmoments} shows results for the first moment of the unpolarized ($\langle x \rangle_{u-d}$) and helicity ($\langle x \rangle_{\Delta u-\Delta d}$) distribution functions. Similar to the previous figures, we include results from all three ensembles, in the same notation as the right panel of Fig.~\ref{fig:gA}. The phenomenological fits of experimental data are shown as horizontal bands. Both plots show a general decrease of the values as $t_s$ increases. However, $\langle x \rangle_{u-d}$ for cA2.90.48 appears to decrease more with $t_s$, especially at $t_s\sim 1.5$. This behavior is likely due to volume rather than quenching effects since the two larger ensembles with different numbers of sea quark flavors, cB211.72.64 and cA2.90.64, are in good agreement. Other than the stronger excited state suppression in the cA2.90.48 $\langle x \rangle_{u-d}$ data, there are no strong volume or quenching effects on these moments. In both plots, the lattice data approaches the phenomenological values as $t_s$ increases, with $\langle x \rangle_{\Delta u-\Delta d}$ overlapping at $t_s\sim 1.6$ but $\langle x \rangle_{u-d}$ is expected to reach agreement at even higher separations.
\begin{figure}
    \hspace*{-.65cm}
    \includegraphics[scale=0.29]{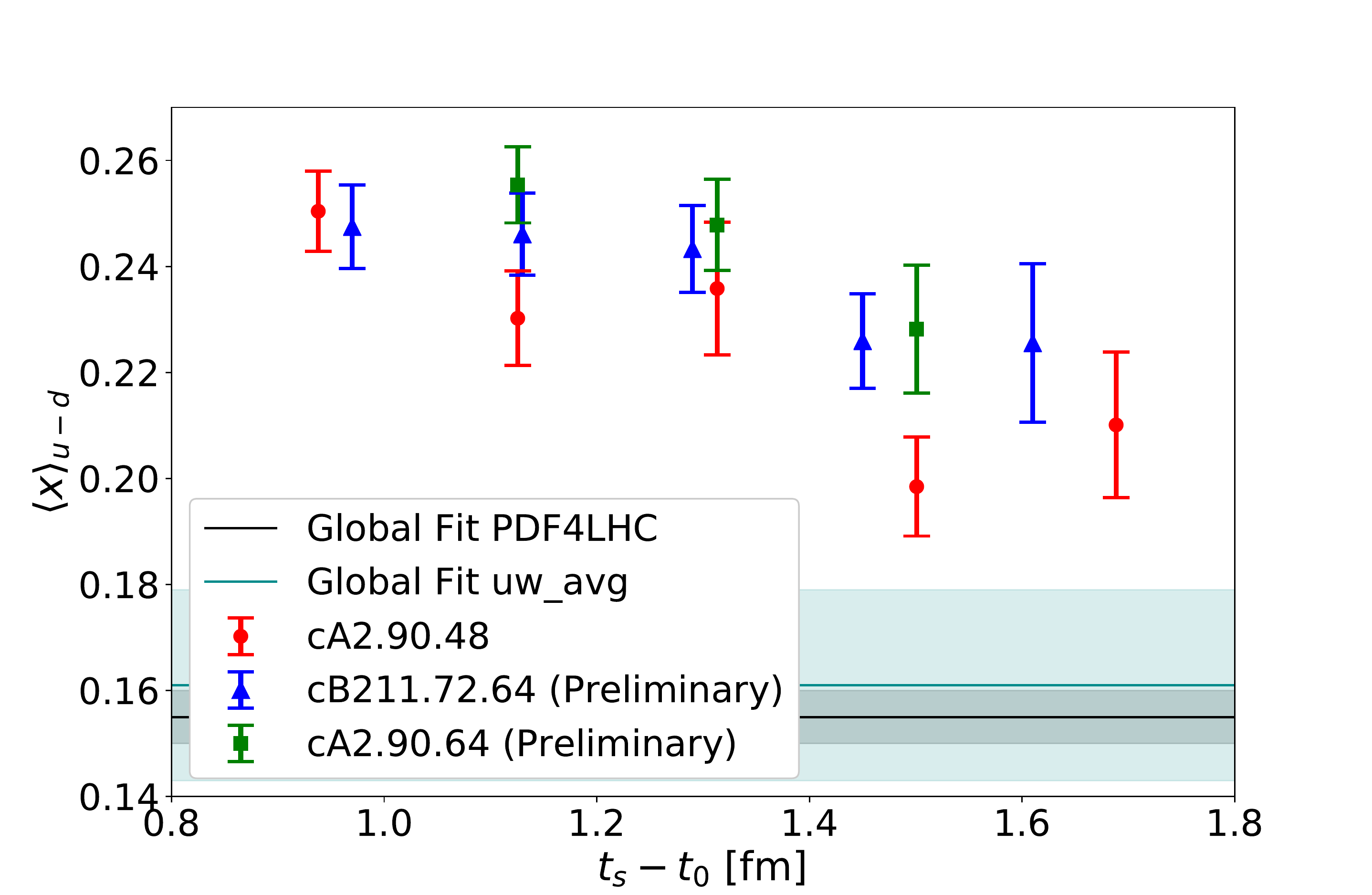}
    \hspace*{-.75cm}
    \includegraphics[scale=0.29]{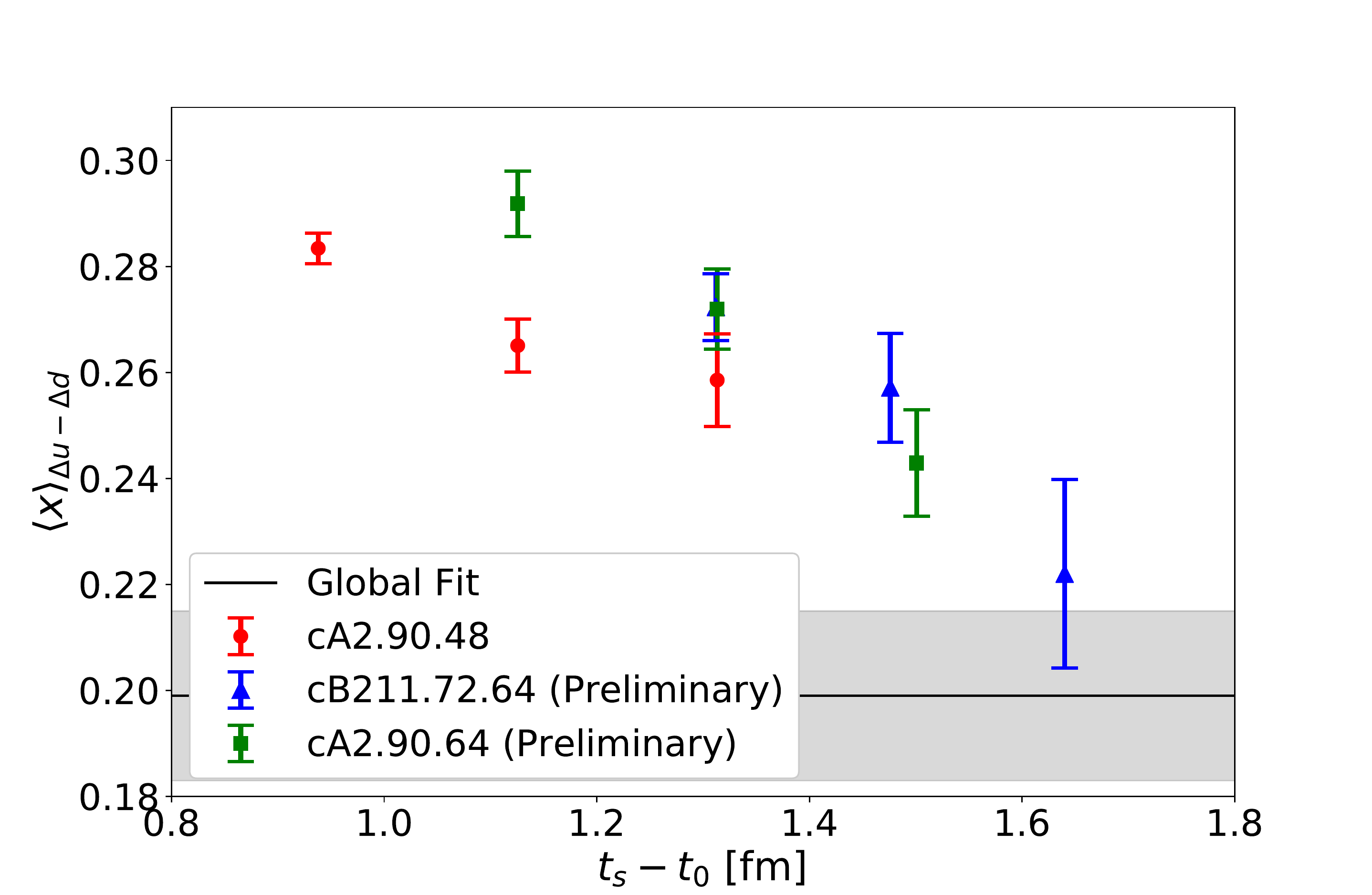}
    \caption{
    Results for $\langle x \rangle_{u-d}$ (left) and $\langle x \rangle_{\Delta u-\Delta d}$ (right). Notation is the same as in the right plot of \ref{fig:gA}. The horizontal bands in the two plots are phenomenological fits of experimental data \cite{wp}.
    }
    \label{fig:PDFmoments}

\end{figure}

\section{Conclusion and Outlook}
In these proceedings we present results on the connected contributions of nucleon charges and the first moments of the unpolarized and helicity PDFs, using two physical pion mass ensembles of varying volumes, one with two dynamical quarks in the sea ($N_f{=}2$) and one also with a strange and a charm in the sea ($N_f{=}2+1+1$). We employ the twisted mass fermions with a clover term and Iwasaki gluons. Results from this work are compared to values obtained by another ensemble of twisted mass fermions that has a smaller volume, with intention a preliminary assessment of volume and quenching effects. Excited states on each ensemble are suppressed by increasing the source-sink separation, seeking convergence between single- and two-state fits. 

We find non-negligible volume and/or quenching effects in $g_A$. In particular, the cA2.90.64 results for these quantities show no apparent decrease in excited state contamination over different source-sink separations. Within our present errors, we cannot clearly resolve finite volume effects for $\langle x \rangle_{u-d}$, but we see a trend towards the phenomenological value for increasing source-sink separation. For the quantities which can be compared to experimental measurements---$g_A$,  $\langle x \rangle_{u-d}$, and  $\langle x \rangle_{\Delta u-\Delta d}$---our results approach the experimental values as the source-sink separation is increased and we find agreement between theory and experiment at larger source-sink separations for $g_A$ and $\langle x \rangle_{\Delta u-\Delta d}$.

Our study will be extended with increase of statistics and analysis for higher values for the source-sink separation, in order to assess and eliminate systematic uncertainties. In the near future we anticipate two more ensembles of $N_f{=}2{+}1{+}1$ twisted mass fermions at the same physical volume as one of the ensembles presented in this work ($N_f{=}2{+}1{+}1$, $64^3{\times}128$) and smaller lattice spacing. These ensembles will be used for an extrapolation to the continuum limit that has never been studied for simulations at the physical point. 

\section*{\it Acknowledgments:}

We want to thank all members of the ETMC for a fruitful collaboration.
This work used computational resources from Extreme Science and
Engineering Discovery Environment (XSEDE),
which is supported by National Science Foundation grant number TG-PHY170022.
The authors gratefully acknowledge the Gauss Centre for Supercomputing e.V. (www.gauss-centre.eu) for funding this project by providing computing time on the GCS Supercomputer SuperMUC at Leibniz Supercomputing Centre (www.lrz.de).
This work was supported by a grant from the Swiss National Supercomputing Centre (CSCS) under project ID s702.
This project has received funding from the European Union's Horizon 2020 research and innovation programme under grant agreement No. 642069. S. B. was supported by this program.
M.C. and C.L. acknowledge financial support by the U.S. National
Science Foundation under Grant No. PHY-1714407.
C.L acknowledges support from the U.S. Department of Energy, Office of Science, Office of Nuclear Physics, contract no. DE-AC02-06CH11357.

\end{document}